# Spatial Tangible User Interfaces
# for Cognitive Assessment and Training

Ehud Sharlin[1], Yuichi Itoh[1], Benjamin Watson[2], Yoshifumi Kitamura[1],
Steve Sutphen[3], Lili Liu[4], Fumio Kishino[1]

[1] Human Interface Engineering Laboratory
Osaka University, 2-1 Yamadaoka, Suita, Osaka 565-0871, Japan
{ehud,itoh,kitamura,kishino}@ist.osaka-u.ac.jp
[2] Department of Computer Science
Northwestern University, 1890 Maple Ave, Evanston, IL 60201 USA
watsonb@cs.nwu.edu
[3] Department of Computing Science
University of Alberta, 221 Athabasca Hall, Edmonton, AB T6G-2E8, Canada
steve@cs.ualberta.ca
4 Department of Occupational Therapy
University of Alberta, 2-64 Corbett Hall, Edmonton, AB T6G-2G4, Canada
lili.liu@ualberta.ca

**Abstract.** This paper discusses Tangible User Interfaces (TUIs) and their potential impact on cognitive assessment and cognitive training. We believe that TUIs, and particularly a subset that we dub spatial TUIs, can extend human computer interaction beyond some of its current limitations. Spatial TUIs exploit human innate spatial and tactile ability in an intuitive and direct manner, affording interaction paradigms that are practically impossible using current interface technology. As proof-of-concept we examine implementations in the field of cognitive assessment and training. In this paper we use Cognitive Cubes, a novel TUI we developed, as an applied test bed for our beliefs, presenting promising experimental results for cognitive assessment of spatial ability, and possibly for training purposes.

## 1 Introduction

We are exploring a new breed of interface between human and computers, spatial Tangible User Interfaces (TUIs). Our research efforts are geared towards demonstrating that spatial TUIs are practical for resolving meaningful real-life problems, while providing considerable benefits over existing solutions and revealing new possibilities that were not viable without them. Spatial TUIs are still in an infant state of development, and can afford only a simple level of expressiveness. The search for meaningful and significant spatial TUIs applications thus becomes quite challenging.
We chose to apply spatial TUIs in the cognitive assessment domain. On one hand we identified a need for spatial, tangible interaction with computers for cognitive assessments. On the other hand, we found that high level of spatial expressiveness (that is, the required level of detail, miniaturization and spatial flexibility) is not required during cognitive assessment interactions and can even be seen as a drawback.

One of the vehicles we developed for our proof-of-concept is Cognitive Cubes—a specialized spatial TUI for cognitive assessment of constructional and spatial ability. The Cognitive Cubes theme follows a simple assessment model: show participants a virtual 3D (Three-Dimensional) prototype and ask them to reconstruct it physically with a spatial TUI. The prototype presented to the participants is an abstract 3D geometrical shape, constructed of generic-looking building blocks. The TUI consists of a set of identical physical building blocks affording 3D construction, much like Lego™ blocks.

The cognitive abilities we measure, namely spatial and constructional abilities, are skills essential for independent living. Their assessment is an important practical and clinical diagnostic tool and is also indispensable in scientific study of cognitive brain functions. Techniques for assessment include asking patients or participants to perform purely cognitive tasks such as mental rotation, as well as constructional tasks involving arrangement of blocks and puzzle pieces into a target configuration. These constructional tasks have the advantage of probing the ability to perceive, plan, and act in the world. Studies suggest that assessment with 3D forms of these tasks may be most demanding and sensitive. However, use of 3D tasks in assessment has been limited by their inherent complexity, which requires considerable examiner training, time and effort if scoring is to be consistent and reliable.

With Cognitive Cubes we demonstrated the first automatic tool for simple, reliable and consistent 3D constructional ability assessment. In this paper we discuss several interdisciplinary aspects of Cognitive Cubes, a fusion of concepts extracted from Human Computer Interaction (HCI) and cognitive assessment. We follow by presenting Cognitive Cubes and detailing the experimental work performed with the system. We emphasize our comparative analysis of Cognitive Cubes and the paper-based Mental Rotation Test (MRT), and the tentative results indicating unexpected improvement in the MRT results after training with Cognitive Cubes.

## 2 Spatial Tangible User Interfaces

HCI research is a multi-disciplinary effort attempting to lower the barriers between people and computers. A key facet of HCI research is directed at the exploitation of our innate tactile and spatial abilities. A successful outcome of this effort is the common mouse. However, human computer interfaces still largely fail to capture the full capacity of our innate ability to manipulate tangible objects. This can be simply demonstrated by comparing the ease of moving, manipulating, assembling and disassembling physical objects with the difficulties of performing similar tasks in a 3D computer-based virtual world using the WIMP (Windows-Icon-Menu-Pointer) interface.

Recently, the notion of a tangible user interface (TUI) has emerged [17], suggesting more elaborate use of physical objects as computer interfaces. Ullmer and Ishii, from the Tangible Media Group at the MIT Media Lab, define TUIs as "devices that give physical form to digital information, employing physical artifacts as representations and controls of the computational data" [17]. TUIs make sense since they engage our natural talents for handling every-day objects in the physical world. We believe TUIs' uniqueness lies in their spatiality; following this line of thought we define ***Spatial TUIs*** as *tangible user interfaces used to mediate interaction with shape, space and structure in the virtual domain*.

Based on previous research, we introduce simple explanations to what makes our tangible-physical interaction seem natural: the support of intuitive spatial mappings, the support of unification of input and output (or I/O unification) and the support of trial-and-error actions.

1. ***Spatial Mapping.*** We define spatial mapping as the relationship between the object's spatial characteristics and the way it is being used. The physical world usually offers clear spatial mapping between objects and their functions (see also [8]). In the digital world most user interaction techniques, particularly in 3D modeling, include a set of rules and controls that manipulate their functionality. However, these rules and controls, implemented with the restrictions of the WIMP interface, are far from enabling an intuitive spatial mapping between interface and application.

2. ***I/O Unification.*** Interaction with physical objects naturally unifies input and output. We see two components to this unification: (1) the clarity of state; the state of physical interfaces in the physical world is usually clearly represented. In HCI this is not a given and the state of the application and the interface do not necessarily mirror each other. (2) the coupling of action and perception space; when we interact with an object in the physical world, our hands and fingers (parts of our action space) coincide, in time and space, with the position of the object we are handling (part of our perception space) [1,2]. This spatial and temporal coupling of perception space and action space focuses attention at one time and place, and enables us to perform complex tasks. Yet the WIMP interaction paradigm separates mouse from screen, input from display, and action from perception. This requires the user to divide her attention, and mentally map one space to the other.

3. ***Support of "Trial-and-error" Actions.*** When we build a physical 3D model, we actually perform an activity that is both cognitive, or goal related, and motorized [1,2]. Such a physical task involves both pragmatic and epistemic actions [1,2,4]. Pragmatic actions can be defined as the straightforward maneuvers we perform in order to bring the 3D shape closer to our cognitive goal. Epistemic (or "trial and error") actions, on the other hand, use the physical setting in order to improve our cognitive understanding of the problem. Some of these epistemic maneuvers will fail and will not bring us any closer to our goal, while others will reveal new information and directions leading to it. In fact, this information might have been very hard to find without trial-and-error [4]. The WIMP interface is geared towards pragmatic actions, with poor support for epistemic actions. For example, the "undo" operation is linear, meaning that to "undo" a single erroneous operation, you have to also "undo" all the operations that followed it.

We employ these three simple observations of the physical world as heuristics for designing effective TUIs.

## 3 On Cognitive Assessment and Technology

When addressing the question of intelligence, Edwin Boring said in 1923 that "intelligence is what the tests test" [10]. Similarly, any assessment of human cognition is shaped, and limited, by the tools it employs. Examining the technological component of cognitive assessment techniques reveals a stagnant state. In 1997 Robert Sternberg highlighted that intelligence testing had changed very little over the last century, making little use of the powerful technology presently available [11,13,16].

Many research endeavors are targeting this deficiency. Since cognitive assessment is all about attempting to have a glimpse of human cognition, state of the art HCI technology should have a dramatic impact on the field. VR, a far-reaching HCI paradigm, is already being exploited as a research test bed for a number of novel cognitive assessments [10-13]. We see TUIs playing an important role in pushing forward the field of cognitive assessment.

Cognitive assessment is a scientific attempt to study cognition and measure human behavior [7]. Testing human behavior involves giving the participant an opportunity to "behave" and measuring it. A measurement tool should be reliable (yielding the same results consistently on different occasions) and valid (measure what it is supposed to measure) [10]. Obviously the measurement tool should be sensitive, safe and should offer the assessor full control over the data collection process [13].

Allowing the participant "to behave" involves the presentation of stimuli which trigger recordable reactions by the participant. Arguably, many classic, paper-pencil cognitive assessment tests offer very limited stimuli, little freedom to behave and low ecological validly (that is, little relevance to normal, everyday human behavior in the real world) [10].

### 3.1 Automation of Cognitive Assessment

Most major psychological paper-pencil tests have been automated or are expected to be automated in the near future [3]. The immediate advantage of this kind of automation is the saving in professional's time: the computer tirelessly samples the participant actions and reliably stores, and refers to a vast assessment knowledge, dramatically reducing the expertise requirements from the assessor. Other obvious advantages of automation are an extremely high density of measurement, an elimination of tester bias and a potential improvement in test reliability. Computerized tests can also be sensitive to response latency, and enable questions tailored based on the examinee's past answers [3]. Automated assessment has also been criticized with concern focused on miscalibration of tests with respect to their written parallels and misuse of tests by unqualified examiners.

We share the view that this kind of straightforward automation portrays merely the tip of the iceberg for automation, and that much of the naysayers' arguments against automation are based on tradition rather than on scientific vision. Robert Sternberg suggested automation-supported "dynamic assessment", where tests targeting learning offer guided performance feedback to the participant [13,16]. Major efforts address the potential of VR for cognitive assessment. VR-based cognitive assessment should afford all the obvious benefits of automation, particularly almost ideal assessment reliability [10,13].

Albert Rizzo and his colleagues (and others) promote VR-based cognitive assessment as a breakthrough in the field, enhancing assessment validity and everyday relevance [9,10,12,13]. VR-based cognitive assessment can "objectively measure behavior in challenging but safe, ecologically valid environments, maintaining experimental control over stimulus delivery and measurement" [13]. VR-based cognitive assessment also introduces many new challenges. One largely unaddressed need is the analysis of huge number of measurements the automated tools extract ("drowning in data" [10]), compared to the simplistic measures of traditional assessment (commonly a single time-to-completion measure per task).

**3.2 Probing Constructional Ability**

Muriel Lezak defines constructional functions as *"perceptual activity that has motor response and a spatial component"* [6]. Constructional ability can be assessed by visuoconstructive, spatial tasks that involve assembling, building and drawing. In a typical constructional assessment, the participant is presented with a spatial pattern and is asked to mimic it by manipulating or assembling physical objects [6]. The test administrator scores participant performance using measures such as time-to-completion and accuracy, or more demanding observations such as order of assembly and strategy analysis. As far as we know, none of these tests were ever automated or computerized.

Constructional functions and disorders can be associated with impairments such as lesion of the non-speech, right-hemisphere of the brain and early phases of AD (Alzheimer Disease), and can be useful in their assessment [3,6]. Constructional function assessment based on the assembly of physical tangible objects generates assessment tools that are non-verbal, relatively culture-free and can be very sensitive to and selective for constructional ability alone [3]. 2D (Two-Dimensional) and 3D constructional tasks have been shown to distinguish between different levels of impairment, suggesting that the more complex 3D construction tasks might be more sensitive to visuoconstructive deficits that were not noticeable on the simpler 2D tasks [6].

2D constructional assessments are widely used. WAIS (Wechsler Adults Intelligence Scale) contains two 2D physical construction subtests, Block Design and Object Assembly [3,6]. In the former, the participant arranges red and white blocks to copy a presented pattern. In Object Assembly, the participant solves a 2D puzzle. Measures for both tests are based on time and accuracy [3,6].

3D constructional assessments are far less common. Two examples are [6]: Block Model from Hecaen et al. and Three Dimensional Constructional Praxis from Benton et al. . In both of these tests the participant tries to match a 3D prototype using wooden blocks, and is scored on time and accuracy. The use of Lego blocks was suggested for 3D tests [6], but to our knowledge was never implemented. Given the complexity of the target shapes in these 3D assessments, manual scoring of even simple measures such as accuracy can be very difficult. Manual scoring of denser measures such as order and strategy would certainly require a very skilled, trained and alert assessor.

### 3.3 Mental Rotation Test

The Mental Rotation Test (MRT) is a paper-pencil based assessment of the visuospatial ability to "turn something over in one's mind" [12,15]. This ability underlies many everyday activities, for example, using a map, or some components of driving [12]. The MRT is 3D and spatial, but in its common form it does not have physical nor constructional components and is purely cognitive.

The MRT is based on early work by Shepard and Metzler [15] that was further established in Vandenberg and Kuse's MRT [18]. MRT's participants are presented with a group of five perspective drawings of 3D objects, one of them is the prototype (the "criterion") object and the rest consists of two identical, but rotated objects, and two "distractor" objects (mirror images of the prototype or simply different objects). The participant is asked to find and mark the two objects that are identical to the prototype object [18].

It was shown that the time needed to determine whether two MRT perspective drawings of objects are similar or not is a linear function of the angular difference between them [15], suggesting that people perform the MRT tasks mentally as if they were physically rotating the objects. The MRT's almost perfect linear relationship between task difficulty and observable human behavior is rare in cognitive assessment; following this relationship the MRT received considerable attention and was extensively researched. The Virtual Reality Spatial Rotation (VRSR) is an automated, VR-based derivative of the MRT [12]. In the VRSR participants are asked to manually orient an MRT-like object until it is superimposed on a target prototype. The VRSR adds a motoric component and enhances the ecological validity of the MRT by presenting the task in a highly immersive VR environment and by enabling the participants to manipulate the virtual object using a tracked physical prop [12].

## 4 Cognitive Cubes

Cognitive Cubes was designed as an automated tool for examination of 3D spatial constructional ability [14]. Cognitive Cubes makes use of ActiveCube [5], a Lego-like tangible user interface for description of 3D shape. With Cognitive Cubes, users attempt to construct a target 3D shape, while each change of shape they make is automatically recorded and scored for assessment.

We created Cognitive Cubes closely following our TUI design heuristics (Section 2). First and foremost, Cognitive Cubes offers a very intuitive spatial mapping between the TUI and the assessment task. Most of the constructional assessment activity is performed entirely in the physical domain, using the physical cube-based TUI which, much like Lego blocks, naturally affords constructional activity. The assessment task involves the presentation of a virtual 3D prototype that the participant attempts to physically reconstruct. We kept the virtual prototype in close visual agreement with the physical cubes, texturing it with a detailed matching texture, sampled from the physical cubes (see Fig. 1 and Fig. 2).

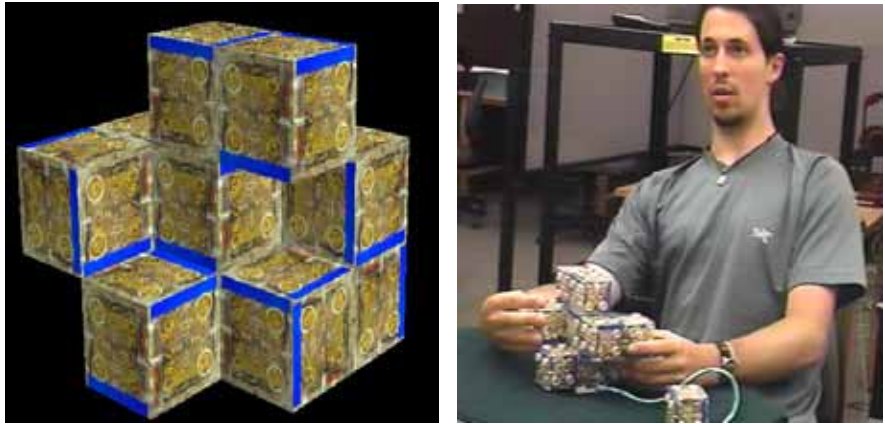

**Fig. 1**. Cognitive Cubes: virtual prototype (left); physical interaction (right)

At first glance, Cognitive Cubes do not offer strong I/O unification because the virtual prototype is presented separately from the physical interface. This argument would have been true if Cognitive Cubes were used for 3D design, however, Cognitive Cubes was used for cognitive assessment. The prototype presented to the participant is merely a visual representation of the cognitive goal the participant is expected to reach, and in this sense the prototype is external to the interaction. A tighter coupling between the presented prototype and the physical TUI would leave very little challenge in the constructional task. We argue that Cognitive Cubes offer good I/O unification since the input—actions performed on the 3D cubes, fully coincide with the output—virtual 3D shapes registered at the host computer.

Lastly, Cognitive Cubes, like many other construction sets, offers extremely flexible exploration of the design domain and trial-and-error actions. Participants can perform actions on the 3D structure in any desired order, undoing their former actions in a completely nonlinear fashion (that is, undoing actions in an order that does not follow the construction order).

As far as we know, Cognitive Cubes is the first computerized tool for assessment of constructional ability, combining the increased sensitivity of 3D constructional tasks with the efficiency, consistency, flexibility and detailed data collection provided by automation.

**4.1 Hardware**

To measure 3D constructional abilities we needed an interface that would maintain a high level of 3D physical constructional expressiveness while enabling precise real-time sensing of the structure's geometry. We chose ActiveCube [5] as the infrastructure to Cognitive Cubes. ActiveCube is arguably the best current example of a spatial 3D TUI for structural input, enabling real-time, interactive, step-by-step, geometry sampling of a 3D structure.

ActiveCube consists of a set of plastic cubes (5 cm/edge) that can be attached to one another using male-female connectors (employing simple clothing-like snaps), forming both a physical shape and a network topology. Each cube and cube face has a unique ID. A host PC is connected to a special base cube and communicates with the small CPUs in each cube through a broadcast mechanism to sense the (dis)connection of any cube. Since all cubes have the same size and shape, any topology represents a unique collective shape [5]. ActiveCube capabilities include more than 3D geometry input. The cubes are equipped with a large variety of input and output devices, supporting flexible interaction paradigms. Some of the cubes are equipped with ultrasonic, optical (visible and IR), tactile, gyroscopic or temperature sensors. Other cubes are equipped with light, audio, motor and vibration actuators [5].

To support our constructional ability assessment paradigm, and to allow us to assess participants with diverse constructional abilities (we were planning to approach young, elderly, and participants with mild Alzheimer disease), Cognitive Cubes hardware had to support the following functions: (i) Allow flexible 3D geometry input by assembly of physical cubes. (ii) Sample the physical 3D cubes structure in real-time. (iii) Allow easy handling of the hardware. Cubes had to be connected to, and disconnected from, each other in a straightforward manner.

To accomplish these requirements, Cognitive Cubes needed only a subset of ActiveCube capabilities, namely the interactive 3D geometry inputting. In this sense, ActiveCube additional input and output capabilities could well be distracting for Cognitive Cubes purposes. We decided to work only with a generic ActiveCube, using cubes with the same color and shape, without any of the extra ActiveCube sensors or actuators.

To ease the connectivity of the cubes we added a blue stripe on each of the cubes faces. To snap the connectors for proper assembly required that the user either match the male-female connectors, or match the two blue stripes on the two connecting faces (see Fig. 1 and Fig. 2 for Cognitive Cubes appearance).

### 4.2 Software

To assess constructional ability using Cognitive Cubes, we needed to present the participant with the prototype she is asked to construct. We chose a virtual display, rather than an actual physical model, as our prototype presentation method. While virtual displays impose a certain level of abstraction (the virtual object is not really there), they can offer relatively high levels of realism and afford an extremely flexible prototype presentation, enabling us to test and edit easily the vocabulary of our prototype shapes. We chose to project the 3D virtual prototype in front of the participant using a monoscopic digital rear projector and a large screen. Viewing all the prototype aspects was achieved by continuously rotating the virtual prototype at a constant rate around its vertical axis, providing 3D depth information. The rotating prototype also engages the participant in mental rotation and use of memory as the virtual proto-type and the physical cubes orientations match only periodically. After several iterations we fixed the prototype rotation at a slow 2.7 revolutions per minute rate.

Other than the display, Cognitive Cubes software supported interaction with minor audio cues: when the participant connects a cube to the structure a distinct chime sounds through a speaker. If the participant chooses to disconnect a cube, a different chime sounds.

During an experiment, the assessor could easily switch between virtual prototypes using a simple menu tool. The software did not generate any cues about the precision of the physical Cognitive Cubes structure vis-à-vis the virtual prototype. Hence the participant worked freely and when satisfied with the match between his construction and the prototype, she informed the assessor, who advanced the system to the next trial. The assessor could also choose to stop the assessment at any point if, for example, the participant was not making any progress.

While the participant attempts to reconstruct the virtual prototype Cognitive Cubes collects a data vector, containing the following values, for each participant action: (i) Event time: in seconds, measured from the time the virtual prototype appeared on the display. (ii) Action type: cube connection or disconnection. (iii) Cube location: can be viewed as a Cartesian set of coordinates, measured from the base-cube which is located at the origin.

After assessment the collected data is analyzed offline to calculate the 3D similarity between the participant's structure $s$ and the prototype $p$. *Similarity* is calculated for each connect or disconnect event. For example, a five step participant assembly will result in five different *similarity* calculations. The equation for *similarity* is presented in Equation (1), where $i$ is an intersection of $s$ and $p$, and $|i|$, $|s|$, and $|p|$ are the number of cubes in $i$, $s$ and $p$. $s$ is maximized over all possible intersections $i$ produced by rotating or translating $s$. Intuitively speaking, *similarity* is the number of intersecting cubes minus the number of remaining "extra" cubes in the participant's structure, normalized by the number of cubes in the prototype.

$$Sim = 100 \cdot \left( \frac{|i|}{|p|} - \frac{|s|-|i|}{|p|} \right) \quad (1)$$

We make the *similarity* at task completion, calculated as described above, one of our four assessment measures. The remaining three are: *last connect*, the time elapsed from the start of the task to the last cube connect or disconnect; *derivative*, the differences between two successively measured similarities in a task divided by the time elapsed between those measurements (local "slope" of the similarity function), averaged for all such pairs in a task; and *zero crossings*, the number of times the local slope crossed zero. We sometimes use the terms "completion time", "rate of progress", and "steadiness of progress" as substitutes for *last connect*, *derivative*, and *zero crossings*.

**4.3 Testing Cognitive Cubes**

We included in our experimental design four task types (Fig. 2). *Intro* tasks were simple practice trials, designed to introduce the participant to Cognitive Cubes. A cube appears on the display after each new connection, indicating the next cube to attach. The *follow* task type also provided step-by-step guidance, but the tasks were much more difficult. *Match* tasks provided no cube-by-cube guidance, but rather displayed a complete virtual prototype for the participant to construct using their own approach. In all three of these task types, the starting point for the participant's construction was the base cube. With *reshape* tasks the participant started from a more complex initial condition (always the same 7-cube 3D construct)—in all other respects *reshape* was exactly like *match*.

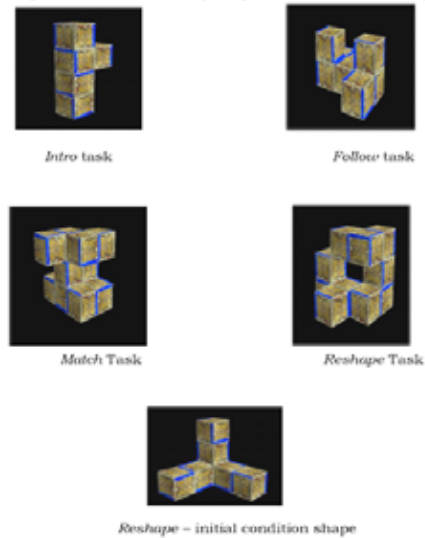

**Fig. 2**. Cognitive Cubes task types (samples)

The participant sat at a table with only Cognitive Cubes placed in front of him. A 125 cm diagonal image was displayed in front of the viewer at a viewing distance of 185 cm using a digital projector, in a brightly lit room (Fig. 1).

The experiment was conducted with a strict written protocol read out loud to the participant. The participant was introduced to the system, the experiment, and its purpose, and then read an information letter. She was told that she might stop the experiment at any time, and asked to sign a consent form. The participant was given a short interview, answering questions concerning age, education, occupation, experience in 3D design, construction sets, computer games, general health and handedness.

**4.4 Results at a glance**

First, we performed an extensive pilot study which included 14 young, healthy participants who performed the entire cognitive assessment. An important lesson learned from the pilot study related to the difficulty of Cognitive Cubes. We found that many of our healthy, young participants faced difficulties with tasks that involved a rela-tively small number of cubes in a 3D arrangement. For example, several of the pilot study participants found a seemingly simple, five-cube *follow* task quite challenging, though eventually manageable. Matching a ten-cube prototype proved to be very challenging for several participants. Consequently, we decided that all prototype shapes would be restricted to at most ten cubes.

To confirm and improve the sensitivity of Cognitive Cubes, we studied its response to three factors known to correspond to differences in cognitive ability: participant *age* (≤34, ≥54), *task type* (*follow*, *match* and *reshape*), and *shape type* (2D, 3D).

Since cognitive ability declines gradually with increasing age, in this study we expected younger participants to perform better than older participants. As the cognitive load of a task increased, cognitive abilities are stressed, leading us to expect better performance with *task types* that required less planning. Similarly, we have already noted the heavier cognitive demands involved in working with 3D shapes. We anticipated better performance with 2D shapes than with 3D shapes.

The cognitive sensitivity study included 16 participants, ranging in age from 24 to 86, with 4 females and 12 males. 7 of the participants were young, 7 elderly and 2 were elderly with mild Alzheimer Disease (AD).

Fig. 3 provides a view of some of the study results. The figure presents the *similarity* (Equation 1) versus time, for a single Cognitive Cubes task. The task is a seven-cube, 3D *match* task. The *similarity* measure curves are plotted for the 13 cognitive sensitivity study participants who performed the task.

It is interesting to note that all the participants who began this task completed it, reaching a final *similarity* of 100%. The *total time* measure varies considerably between the different groups: most of the young participants completed the task faster than most of the elderly participants. All of the participants accomplished the task more quickly than the single AD participant.

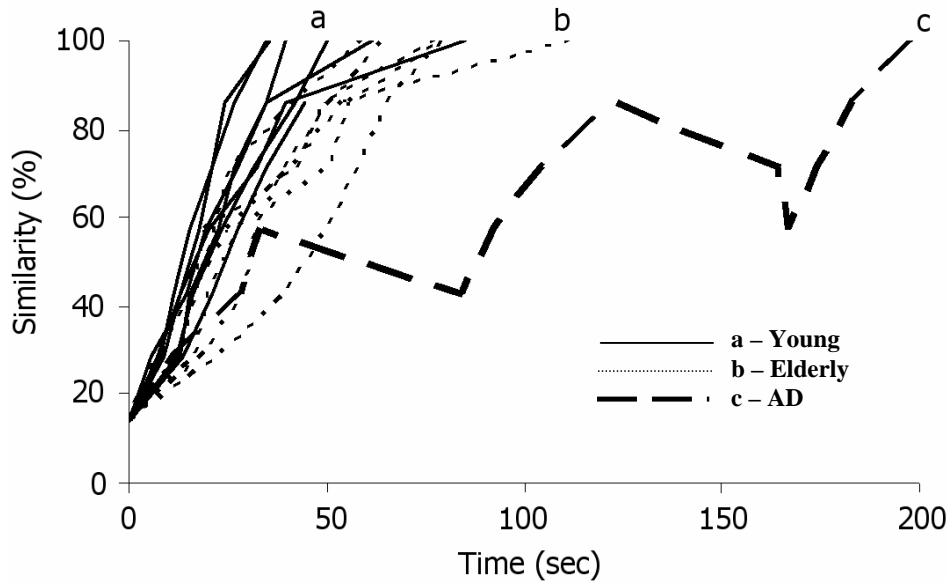

**Fig. 3.** Similarity vs. time; cognitive sensitivity study; single task

Participants' rate of progress, as manifested through the curve slope, or the *derivative* measure, also differs between the groups. Most of the young participants have a steeper similarity slope than the elderly participants. All the participants have a faster rate of progress than the AD participant.

Lastly, with the exception of the AD participant, all participants have steady progress towards the goal, and thus no *zero crossings* (Section 4.2). However, the AD participant similarity curve local slope crosses zero several times.

We analyzed the results using one 3-factor (2 *age* x 3 *task type* x 2 *shape type*) ANOVA for each of the *last connect*, *similarity*, *zero crossings*, and *derivative measures*. Because there were so few AD participants, we excluded them from any analyses of variance. We also exclude the *intro task type* from the analyses.

All three factors produced main effects in line with our expectations. Participant performance varies significantly by age, with elderly participants needing more time to complete each task, and showing a low rate of progress. By all four dependent measures, participant performance is also significantly affected by *shape type*. 2D shape construction is completed more quickly, more accurately, and with a higher and steadier rate of progress. Finally, *task type* also has significant effects on all four measures. *Follow* is the easiest of the task types, enabling quick completion and a high, steady rate of progress toward the target shape. However, shape *similarity* is lowest with the *follow* task. Participants perform the *match* and *reshape* tasks with roughly equal completion times and similarities, but the rate of progress in the *match* task type is higher and steadier. A thorough discussion of our cognitive sensitivity study results is left out of the scope of this paper, and is presented elsewhere [14].

## 5 Cognitive Cubes and the Mental Rotation test (MRT)

Having studied the sensitivity of Cognitive Cubes to factors related to cognitive performance, we turn to a comparison of Cognitive Cubes to a known tool for 3D spatial assessment: the MRT. As we discussed in Section 3.3 the MRT has 3D and spatial components, like Cognitive Cubes, leading us to expect a strong relationship between the two assessments, particularly with 3D tasks. However, since the MRT does not include any of Cognitive Cubes' constructional, planning, or motor task components, we might anticipate the relationship to be limited to simpler tasks such as *follow*.

### 5.1 Participants and Methods

The test comparison study's 12 participants had ages ranging from 18-36, with an average age of 27.66 and standard deviation of 5.61 years. Four of the participants were females and eight were males. Participants were all volunteers recruited on and off campus none of whom participated in any other phases of the Cognitive Cubes experiments. The procedure followed the general Cognitive Cubes experimental methodology except that participants took the MRT test before and after the Cogni-tive Cubes assessment. We refer to these MRT sessions as Pre-Cognitive Cubes, and Post-Cognitive Cubes, respectively.

**5.2 Results**

We start this section with a few selected comments from Cognitive Cubes participants, based on post-session interviews. We tried to include negative feedback as well as positive, though most experiences were very positive.
Male, healthy, 34—"Cognitive Cubes were easier and more fun than the MRT, since they are tactile there's less to think about".
Female, healthy, 28—"MRT is more challenging than Cognitive Cubes. With Cognitive cubes you can always 'try', with the MRT you have to use only your imagination".
Female, healthy, 29—"MRT is stressful, feels like an exam. Cognitive Cubes is fun, you feel you are doing something. Less stress and there's no time limit".

We performed our MRT/Cognitive Cubes comparisons using correlations. As will be discussed later in this Section, our post Cognitive Cubes MRT reached ceiling, meaning that most of the subjects scored very high MRT scores after their Cognitive Cubes session. Because they reached ceiling and lost sensitivity, correlating of post-Cognitive Cubes MRT to Cognitive Cubes measures would be a meaningless exercise and are not presented.
The correlations of pre-Cognitive Cubes MRT and Cognitive Cubes are presented in Table 1, along with the probability that the correlations are not significantly different from 0. Those correlations with high probability of being different from 0 are presented in bold, underlined digits. The measure with the most significant overall correlation (and the only reaching marginal significance) is the *derivative* measure. Correlations to *zero crossings* are low. Correlations to *similarity* are also low, perhaps because similarities are uniformly high. Correlations to *last connect* are also high. Correlations are only slightly stronger for 3D than 2D shapes, while correlations are strongest with *follow* tasks, slightly weaker with *match* tasks, and completely untrustworthy with *reshape* tasks.
Contrary to our expectations, both 2D and 3D task types produce some good correlations to the 3D MRT. We believe this may well be attributable to task difficulty. While the MRT asks the user to perform a small set of relatively simple 3D mental rotations, Cognitive Cubes challenges participants to construct a single shape, which may be small or large, 2D or 3D. Which is more like the MRT: building from scratch a complex 3D shape, or a simple 2D shape? The answer is unclear, and thus the lack of clarity in the *shape type* correlations.

**Table 1.** Pre-CC MRT/Cognitive Cubes correlations. Correlation significance: (p<.1) in bold, (p<.05) underlined

| Dependent Measures | Overall | Shape Type | | Task Type | | |
|---|---|---|---|---|---|---|
| | | 2D | 3D | follow | match | reshape |
| Last connect | -0.38 | **-0.49** | -0.35 | **-0.63** | -0.35 | -0.24 |
| Similarity | 0.03 | -0.36 | 0.17 | 0.16 | -0.09 | 0.08 |
| Zero crossings | -0.23 | 0.07 | -0.25 | -0.14 | -0.45 | 0.11 |
| Derivative | **0.51** | 0.38 | **0.57** | 0.43 | **0.50** | 0.34 |

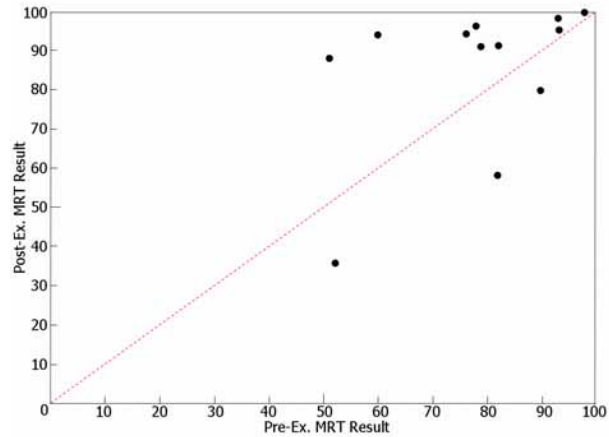

**Fig. 4.** Post-Cognitive Cubes vs. pre-Cognitive Cubes MRT Scores

### 5.3 Surprises

We compared MRT results obtained before (pre-Cognitive Cubes) and after (post-Cognitive Cubes) the Cognitive Cubes assessment. Interestingly, post-Cognitive Cubes MRTs are markedly improved (most in the 90th percentile). The improvement in the MRT score after the Cognitive Cubes sessions is illustrated in Fig. 4. Each graph point represents MRT results from a single participant, its X axis value the pre-Cognitive Cubes MRT score, and its Y axis value the post-Cognitive Cubes MRT score.
While it is well known that repeating the MRT brings improved performance, improvements are in this case well above the normally reported repetition improvement rate of roughly 5%. Although these results are preliminary we believe they indicate some of the potential TUIs have as intuitive and automatic training aids.

## 6 Conclusion

Is Cognitive Cubes a useful tool for assessment? Our experience certainly indicates great promise. Despite being a prototype, the ActiveCube hardware component stood up well to intense use and proved to be quite intuitive for our participants. In experimental evaluation, the system as a whole was sensitive to well-known cognitive factors and compared favorably to an existing assessment. Automation introduced a previously unachievable level of reliability and resolution in 3D measurement and scoring. Despite all this, Cognitive Cubes is not yet ready for regular use.

How might Cognitive Cubes be prepared for use in the field? The gap between a good prototype and a reliable tool is a large one. Use in clinical or research settings would require significant improvements in cost, reduction of connection and system errors, and improvements in structural strength. These are fairly typical requirements for the development of any technology. In addition, extensive testing would be required to identify the distribution of scores typically achieved with Cognitive Cubes. In this way, assessors can reliably decide whether or not a score indicates impairment. One the most unique strengths of Cognitive Cubes is its ability to capture each step of the task progress—closely mirroring the cognitive processing of the participant. With the same data used to build similarity graphs it is also possible to build decision trees reflecting the participant's chosen path through the space of possible cube-by-cube construction sequences. This dynamic process can be probed even more deeply by attempting to categorize participant trees according to cognitive ability.

The improvement from pre-Cognitive Cubes MRT to post-Cognitive Cubes MRT results is unexpected, but very intriguing. Could Cognitive Cubes be used as a form of cognitive therapy or training, for example in rehabilitation? Our current results are preliminary, but we see a great promise for further probing in this direction.

Cognitive Cubes, to our knowledge, is the first system for the automated assessment of 3D spatial and constructional ability. Cognitive Cubes makes use of ActiveCube, a 3D spatial TUI, for describing 3D shape. Cognitive Cubes offers improved sensitivity and reliability in assessment of cognitive ability and ultimately, reduced cost. Our experimental evaluation with 43 participants confirms the sensitivity and reliability of the system.

We see Cognitive Cubes as a proof-of-concept demonstrating our research goal, showing that a specialized spatial TUI closely tied to an application can offer substantial benefits over existing solutions and suggests completely new methodologies for approaching the applied problem.

We also see Cognitive Cubes as a practical and successful example of our TUI design heuristics being put to work. Our choices during the design of Cognitive Cubes, for example, selecting a spatial application domain, and choosing a very intuitive spatial mapping between the interface and the task, were all closely guided by our TUI heuristics. We believe that the success of Cognitive Cubes should also be attributed to the heuristics that guided the design.

## Acknowledgments

We are very grateful for the ongoing help and support of Dr. Jonathan Schaeffer from the University of Alberta, Dr. Albert "Skip" Rizzo from the University of Southern California, and Dr. Alinda Friedman from the University of Alberta. We thank all the members (current and past) of the ActiveCube research group at the Human Interface Engineering Laboratory, Osaka University. This research was supported in part by "The 21st Century Center of Excellence Program" of the Ministry of Education, Culture, Sports, Science and Technology, Japan.